\documentclass[sigconf]{acmart}
\AtBeginDocument{%
  }

\usepackage{tabularx}
\usepackage{soul}
\usepackage{natbib} 
\usepackage{dirtytalk}
\usepackage{fancybox} 
\newcommand{\promptbox}[1]{
  \begin{center} 
    \doublebox{%
      \begin{minipage}{.95\columnwidth} 
        \vspace{5pt} 
        #1
        \vspace{5pt} 
      \end{minipage}%
    }
  \end{center}
}

\copyrightyear{2026}
\acmYear{2026}
\setcopyright{cc}
\setcctype{by-nc-nd}
\acmConference[CHIWORK '26]{Proceedings of the 5th Annual Symposium on Human-Computer Interaction for Work}{June 22--25, 2026}{Linz, Austria}
\acmBooktitle{Proceedings of the 5th Annual Symposium on Human-Computer Interaction for Work (CHIWORK '26), June 22--25, 2026, Linz, Austria}
\acmDOI{10.1145/3808045.3808093}
\acmISBN{979-8-4007-2429-9/2026/06}

\begin{document}

\title[MultEval]{MultEval: Supporting Collaborative Alignment for LLM-as-a-Judge Evaluation Criteria}

\author{Charles Chiang}
\email{cchiang3@nd.edu}
\orcid{0009-0008-6079-4355}
\affiliation{%
  \institution{University of Notre Dame}
  \city{Notre Dame}
  \state{IN}
  \country{USA}
}

\author{Simret Gebreegziabher}
\email{sgebreeg@nd.edu}
\orcid{0000-0002-1772-6065}
\affiliation{%
  \institution{University of Notre Dame}
  \city{Notre Dame}
  \state{IN}
  \country{USA}
}
\author{Annalisa Szymanski}
\email{aszyman2@nd.edu}
\orcid{0009-0009-5472-282X}
\affiliation{%
  \institution{University of Notre Dame}
  \city{Notre Dame}
  \state{IN}
  \country{USA}
}
\author{Yukun Yang}
\email{yyang35@nd.edu}
\orcid{0009-0003-1971-4468}
\affiliation{%
  \institution{University of Notre Dame}
  \city{Notre Dame}
  \state{IN}
  \country{USA}
}
\author{Hyo Jin Do}
\email{dohyojin90@gmail.com}
\orcid{0000-0001-8297-6792}
\affiliation{%
  \institution{IBM Research}
  \city{Yorktown Heights}
  \state{NY}
  \country{USA}
}
\author{Zahra Ashktorab}
\email{zashktorab@gmail.com}
\orcid{0000-0002-0686-7911}
\affiliation{%
  \institution{IBM Research}
  \city{Yorktown Heights}
  \state{NY}
  \country{USA}
}
\author{Werner Geyer}
\authornote{This work was done in connection with IBM Research. Author is now at Oracle.}
\email{werner.geyer@gmail.com}
\orcid{0000-0003-4699-5026}
\affiliation{%
  \institution{IBM Research}
  \city{Yorktown Heights}
  \state{NY}
  \country{USA}
}
\author{Toby Li}
\email{toby.j.li@nd.edu}
\orcid{0000-0001-7902-7625}
\affiliation{%
  \institution{University of Notre Dame}
  \city{Notre Dame}
  \state{IN}
  \country{USA}
}
\author{Diego Gomez-Zara}
\email{dgomezara@nd.edu}
\orcid{}
\affiliation{%
  \institution{University of Notre Dame}
  \city{Notre Dame}
  \state{IN}
  \country{USA}
}

\renewcommand{\shortauthors}{Chiang et al.}
\newcommand{\toolname}{\textsc{\textbf{MultEval}}} 

\begin{abstract}
  LLM-as-a-judge approaches have emerged as a scalable solution for evaluating model behaviors, yet they rely on evaluation criteria often created by a single individual, embedding that person's assumptions, priorities, and interpretive lens. In practice, defining such criteria is a collaborative and contested process involving multiple stakeholders with different values, interpretations, and priorities---an aspect largely unsupported by existing tools. To examine this problem in depth, we present a formative study examining how stakeholders collaboratively create, negotiate, and refine evaluation criteria for LLM-as-a-judge systems. Our findings reveal challenges in human oversight, including difficulties in establishing shared understanding, aligning values across stakeholders with different expertise and priorities, and translating nuanced human judgments into criteria that are interpretable and actionable for LLM judges. Based on these insights, we developed \toolname{}, a system that supports collaborative criteria by enabling multiple evaluators to surface and diagnose disagreements using consensus-building theory, iteratively revise criteria with attached examples and proposal history, and maintain transparency over how judgments are encoded into an automated evaluator. We further report a case study in which a team of domain experts used \toolname{} to collaboratively author criteria, illustrating how coordination and collaborative consensus-making shape criteria evolution.
\end{abstract}



\maketitle

\section{Introduction}
As Large Language Models (LLMs) are increasingly deployed across different domains, developing robust mechanisms for oversight and evaluation has become a central concern. LLM-as-a-judge has emerged as a scalable alternative to human evaluation, in which LLMs themselves evaluate model outputs according to explicitly defined criteria \cite{Gu_Jiang_Shi_Tan_Zhai_Xu_Li_Shen_Ma_Liu_et, zheng2023judging}. Prior work has shown that strong LLM-judges can achieve moderate to high agreement with human judgments across tasks such as ranking responses and summarization scoring~\cite{dubois2024length}. However, most existing systems assume that a single evaluator defines and refines evaluation criteria \cite{gebreegziabher2025metricmate,kim2024evallm,Ashktorab_Desmond_Pan_Johnson_Cooper_Daly_Nair_Pedapati_Do_Geyer_Aligning_2025,Kim_Lee_Lee_Seering_Kim_2025}. 

\begin{figure*}
  \centering
  \includegraphics[width=\textwidth]{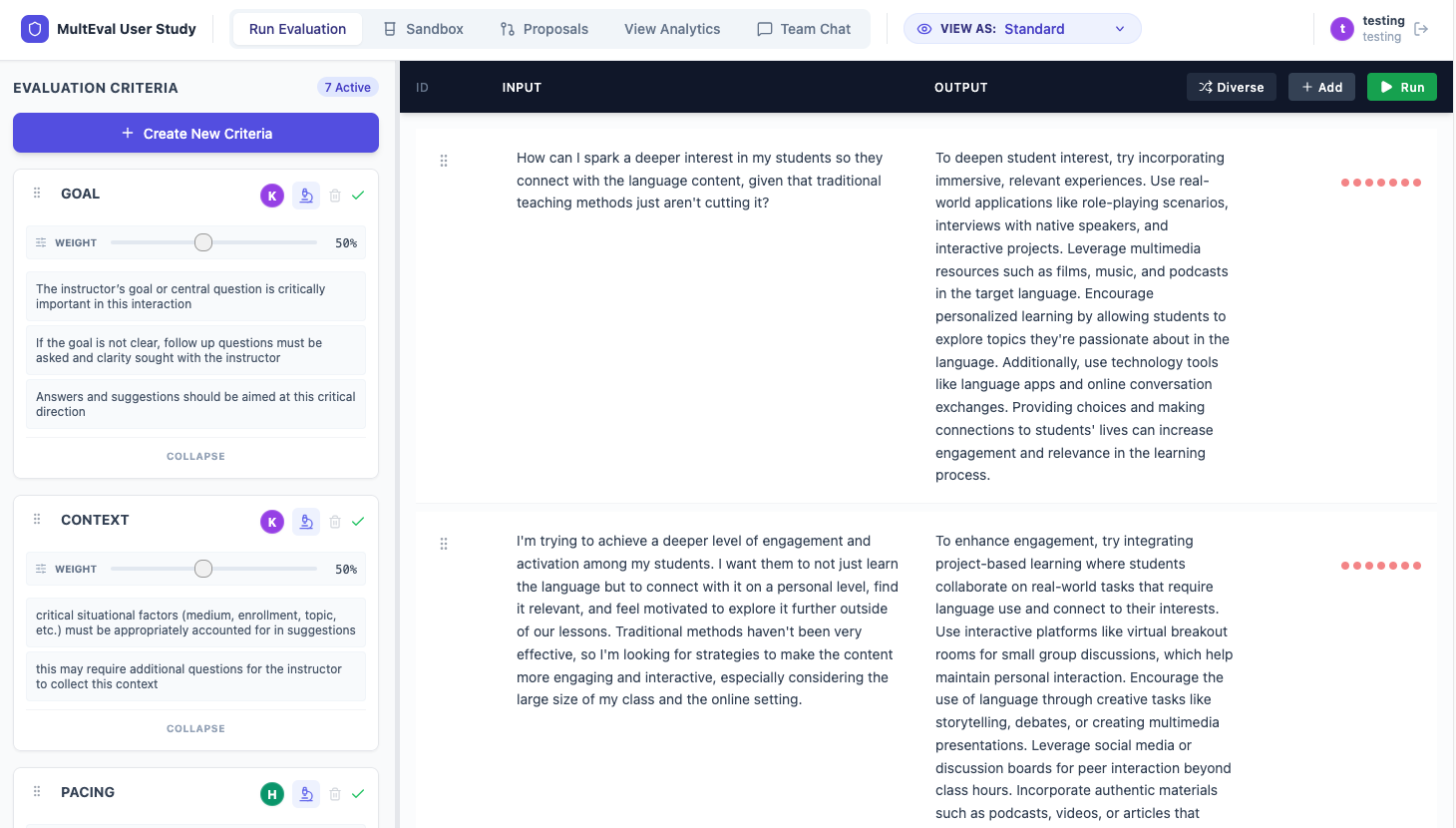}
  \caption{
    The initial evaluation screen. Criteria and assertions are shown on the left, while the dataset is shown on the right. At the top, users can switch between the different tabs. When the evaluation runs, the dots on the right-hand side turn green or red, depending on whether the criteria pass or fail.
  }
  \label{fig:teaser}
\end{figure*}

In practice, defining evaluation criteria is not a neutral or individual task. Rather, it is a collaborative and socially negotiated process involving multiple stakeholders with diverse expertise, values, and interpretations \cite{parmar2010stakeholder,soltani2015multiple}. Teams must establish shared understanding, reconcile competing interpretations, and translate nuanced human judgments into criteria that are both interpretable and actionable for LLM judges. Despite this, current tools for LLM-as-a-judge workflows offer limited support for making such disagreements visible, negotiating competing perspectives, or coordinating consensus around evaluation criteria. This creates a bottleneck in which perspectives remain implicit or unconsidered, and where relevant domain expertise or stakeholder viewpoints are excluded \cite{passi2019problem,friedman2013value}.

To address this gap, we first conducted a formative study to characterize how stakeholders collaboratively create and negotiate evaluation criteria for LLM-as-a-judge systems, identifying key challenges in coordination, alignment, and criteria specification practices. Based on these insights, we introduce \toolname{}, a system designed to support collaborative criteria creation for such systems. Grounded in consensus-building theory \cite{briggs2005toward}, \toolname{} enables multiple evaluators to propose, test, and revise evaluation criteria while surfacing disagreements, diagnosing sources of conflict, and preserving rationale over time. 

We evaluate \toolname{} through a case study in which a real-world team of domain experts collaboratively authored evaluation criteria for a pedagogical LLM, allowing us to examine how teams coordinate, negotiate specificity, and manage roles and authority during criteria creation. Our findings show that collaborative criteria development involves substantial sense-making and coordination work - such as establishing shared vocabulary, translating nuanced judgments into executable criteria, and negotiating decision rights - and that \toolname{}’s proposal-based workflow and transparency features help make these processes more visible and manageable.

This paper makes three contributions. First, we present a formative qualitative study characterizing how multiple stakeholders collaboratively create and negotiate evaluation criteria for LLM-as-a-judge systems, revealing coordination challenges related to value alignment, role asymmetries, and translating human judgment into executable criteria. Second, we derive design goals for supporting multi-stakeholder criteria creation grounded in consensus-building theory. Third, we introduce \toolname{}, a multi-user web system that operationalizes these goals by enabling evaluators to surface disagreement, preserve rationale, and iteratively refine shared criteria. We further report a case study illustrating how these mechanisms shape collaborative criteria development in practice. In sum, this paper provides a new perspective on LLM-as-a-judge workflows by showing that evaluation criteria are not merely technical specifications but socially negotiated artifacts shaped by multiple stakeholders. The system and materials are available at \url{https://github.com/RINGZ-Lab/MultEval}.

\section{Related Work}

\subsection{Interactive Systems for LLM-as-a-judge}
LLM-as-a-judge refers to the use of LLMs as automated evaluators that assess other model outputs according to explicitly defined criteria or prompts \cite{zheng2023judging,dubois2024length}. This approach offers a scalable alternative to human evaluation, with prior work showing moderate to high agreement with human judgments. In this paradigm, evaluation is often framed as aligning model judgments with human preferences, where users iteratively refine criteria and prompts to better approximate human evaluation \cite{shankar2024validates}.

A growing ecosystem of computational applications has emerged to support LLM-as-a-judge workflows. Systems such as \textit{EvalLM} \cite{kim2024evallm} or \textit{Chainforge} \cite{arawjo2024chainforge} allow users to iterate and optimize prompts for an LLM-judge by comparing outputs. \textit{ChatEval} takes a multi-agent approach to iterating LLM judges \cite{Chan_Chen_Su_Yu_Xue_Zhang_Fu_Liu_2023}. Other systems focus more explicitly on criteria creation and refinement. For example, \textit{Promptfoo} \cite{Promptfoo}, \textit{Stax} \cite{Stax_Stax}, and \textit{Chainforge} \cite{arawjo2024chainforge} support users in defining and testing evaluation criteria across different models. \textit{EvaluLLM} takes this a step further, allowing users to run evaluations with different prompts and criteria metrics \cite{Desmond_Ashktorab_Pan_Dugan_Johnson_EvaluLLM_2024}. Systems such as \textit{EvalGen} generate criteria automatically from a history of criterion generation \cite{shankar2024validates}, while \textit{MetricMate} \cite{gebreegziabher2025metricmate} and \textit{EvalAssist} \cite{Ashktorab_Desmond_Pan_Johnson_Cooper_Daly_Nair_Pedapati_Do_Geyer_Aligning_2025} support users in the whole process of criteria generation, iteration, and testing. Lastly, \textit{Evalet} breaks the criteria down into fragments that highlight different aspects of an output depending on its relevance to each criterion \cite{Kim_Lee_Lee_Seering_Kim_2025}. This diversity of systems reflects the growing need to support the iterative development of evaluation criteria \cite{Gu_Jiang_Shi_Tan_Zhai_Xu_Li_Shen_Ma_Liu_et}.

However, all these systems are designed for a single evaluator. In real-world scenarios, these criteria are often created through a collaborative process involving multiple stakeholders with different perspectives and expertise. As a result, different evaluators may interpret the criteria differently and bring different perspectives and requirements to the criteria definition, given their role. Current tools provide limited support for making such disagreements visible or coordinating alignment across evaluators. To address this gap, \toolname{} draws on consensus-building theory \cite{briggs2005toward} to support collaborative alignment by surfacing disagreements and enabling structured negotiation of evaluation criteria among stakeholders.

\subsection{Designing Systems for User Agreement}
While working in a diverse team can increase creativity and performance, it can also introduce disagreements over priorities and interpretations \cite{gomezzara2020,jehn1999differences}. Prior work has examined how interfaces can surface disagreement and support convergence, particularly in civic and political contexts. For example, systems that present news to politically diverse audiences study how to expose users to differing viewpoints without immediately triggering disengagement \cite{Graells-Garrido_Lalmas_Baeza-Yates_2016, munson2010presenting}. Related work also considers large-scale deliberation platforms that let participants collectively propose and vote on policy ideas (e.g., \textit{pol.is} and \textit{vTaiwan}) \cite{hsiao2018vtaiwan}. One effective mechanism shown is introducing ``intermediary'' topics that can connect polarized groups \cite{Graells-Garrido_Lalmas_Baeza-Yates_2016}. Beyond content, collaborative sequencing highlights how the ordering of contributions can shape group convergence over time \cite{kim2021supporting}.
  
Process-oriented perspectives on consensus building further emphasize that agreement is often achieved through structured cycles of proposal, critique, and refinement. For instance, \citet{briggs2005toward} describes a process model of consensus building that characterizes the activities groups undertake when moving from divergent to convergent positions. The Delphi method similarly structures convergence through repeated rounds of elicitation and feedback. Participants answer the same question, review aggregated responses, and revise their answers across rounds until responses stabilize \cite{Wilson_Brazier_Gkatzia_Robertson_2024}. While effective in expert settings, Delphi-style workflows can be costly due to repeated, coordinated rounds of participation \cite{Wilson_Brazier_Gkatzia_Robertson_2024}.

Similar approaches appear in collaborative analytics work. For example, \textit{CollabCoder} supports qualitative coding, in which multiple analysts independently label data and then discuss discrepancies to reach a codebook \cite{Gao_Guo_Lim_Zhang_Zhang_Li_Perrault_2024}. Similar to criteria iteration for LLM evaluation, qualitative codes evolve through discussion and negotiation, and tools can support convergence by making differences visible and helping teams reconcile diverse interpretations \cite{Gao_Guo_Lim_Zhang_Zhang_Li_Perrault_2024}. 

These studies show that agreement is not simply about aggregating votes, but about supporting cycles of independent reflection, disagreements, and convergence on shared artifacts. However, prior deliberation and consensus systems are not tailored to LLM evaluation, where teams must translate discussion into executable, testable criteria. To address this gap, \toolname{} scaffolds a structured proposal-and-review loop around criteria edits, pairs deliberation with concrete evidence (e.g., representative data points and changed evaluation outcomes), and helps teams diagnose disagreement to support more efficient convergence.

\subsection{Multi-stakeholder Human-AI Alignment}
A central challenge in human-AI alignment is deciding \textit{whose} preferences and values should be reflected in the aligned system \cite{gabriel2020artificial,zhi2025beyond}. While many alignment workflows assume a one-to-one relationship between a user and an AI system, emerging work emphasizes settings in which alignment must account for interactions among multiple humans and AI agents. For example, jury learning explores how to learn from panels of annotators and how to surface and manage disagreement \cite{gordon2022jury}. Similarly, collective approaches explore how groups can author and iteratively refine shared principles that guide model behavior~\cite{huang2024collective}. 

This shift toward multi-stakeholder alignment also motivates the design of interactive systems that support the articulation, inspection, and negotiation of alignment targets. Work on interactive AI alignment characterizes this paradigm and highlights key challenges, including making trade-offs legible, supporting iteration, and handling conflicting preferences \cite{Terry_Kulkarni_Wattenberg_Dixon_Morris_2024}. In the context of LLM-as-a-judge, incorporating domain experts is critical for producing criteria that reflect real task requirements, as experts and lay users may propose meaningfully different criteria \cite{Szymanski_Gebreegziabher_Anuyah_Metoyer_Li_2024}.

More broadly, alignment has also been conceptualized as the calibration of model behavior to shared values and normative principles. Constitutional AI operationalizes this idea by training models to follow an explicit set of normative rules \cite{bai2022constitutional}, and subsequent work has explored broadening these principles through collective input \cite{huang2024collective}. These approaches highlight a growing interest in multi-stakeholder alignment, where model behavior is shaped by inputs from multiple collaborators \cite{lam2025policy,feng2026}. 

However, much of this work focuses on aggregating or encoding preferences, with less emphasis on how stakeholders collaboratively articulate, negotiate, and refine alignment targets in practice. Tools that support interaction, interpretation, and coordination play a key role in enabling collaborative alignment by helping stakeholders make sense of model outputs and reconcile differing interpretations \cite{Gero_Swoopes_Gu_Kummerfeld_Glassman_2024}. \toolname{} addresses this gap by supporting both human–human alignment and human–AI alignment within a shared workflow for collaborative criteria creation.

\section{Formative Study}
Grounded in our literature review, we conducted a formative qualitative study using semi-structured interviews with 10 participants who had experience working on LLM-based systems and participating in LLM evaluation efforts. The goal of this study was to understand how different teams create evaluation criteria, with particular attention to how individuals in different roles contribute to, coordinate around, and make decisions during the evaluation process. These insights inform the design of tools and workflows to better support LLM evaluation across stakeholders. 

Interviews lasted approximately one hour and followed a semi-structured protocol. Prior to the interview, participants completed a short pre-study survey collecting demographic information and background experience. During the interviews, participants were asked to describe their role in LLM evaluation, how they interacted with other stakeholders (e.g., developers, domain experts, end users), and how evaluation activities were conducted over time. Interview questions probed topics such as coordination across roles, iteration practices, communication strategies, conflicts or breakdowns in the evaluation process, and how requirements from different stakeholders were managed. This study protocol was approved by the University of Notre Dame's Institutional Review Board (IRB).

\subsection{Participants}
Participants were recruited through a combination of social media and professional networks (i.e., LinkedIn, Slack), word of mouth, and the authors' institution newsletters. Recruitment material asked for participants with experience working on a team to evaluate LLM outputs, and we pre-screened participants to ensure they had sufficient experience collaborating with others in evaluation contexts and to capture a diverse set of participant roles and perspectives. 

The final sample included 10 participants, with four from industry and six from academia. Participants occupied a range of roles spanning both technical and knowledge-based responsibilities in LLM evaluation, including frontend and backend developers, pedagogy domain experts, and hybrid roles combining development and domain expertise. Technical participants were primarily involved in modeling and training LLMs, testing and quality assurance, and end-user testing. Domain experts contributed through evaluation criteria development, design feedback, testing, and data provisioning. Several participants held multiple roles within their teams, reflecting the interdisciplinary and collaborative nature of LLM evaluation work. Five participants were recruited from the same project, where they worked as domain-knowledge evaluators for a pedagogical chatbot.
Additional details are provided in Table~\ref{tab:participants}.

In describing participant roles, we distinguish between \textit{developers} and \textit{domain experts}. Developers are primarily responsible for building and modifying the underlying LLM system, including model development, prompting, and evaluation infrastructure. Domain experts contribute task-specific knowledge and are typically responsible for defining, interpreting, and assessing evaluation criteria. During interviews, participants often referred to these roles when describing collaboration, including their interactions with counterparts across technical and domain boundaries. While some participants occupied hybrid roles, this distinction helps characterize recurring patterns in how responsibilities and decision-making were distributed across teams.

\subsection{Qualitative Analysis}
All interviews were conducted and recorded via Zoom and transcribed using online transcription tools. We analyzed the interview transcripts using thematic analysis~\cite{braun2006using}. Three researchers independently reviewed the transcripts, first engaging in close reading to familiarize themselves with the data and recording notes. Each researcher then generated initial codes to capture recurring patterns and ideas across the dataset. The research team met to discuss and reconcile these codes, after which the codes were iteratively grouped into candidate themes. These themes were reviewed, refined, and synthesized into higher-level key insights.

\subsection{Key Insights}
\label{formative:keyinsights}

\subsubsection{KI1: Collaborative criteria generation highlights divisions} \label{KI1}
We found that developers and domain experts often work closely together in interdisciplinary teams. While developers drive system development, domain experts are called upon to evaluate the system's outputs and progress. Even though evaluation is a critical and time-consuming iterative process, domain expert participation is often limited \cite{shankar2024validates}. Domain experts were not always involved in day-to-day progress, as many developers stated they needed time to develop the system to a satisfactory state before calling on the domain experts for evaluation. Additionally, domain experts may need technical guidance to understand how to contribute meaningfully. P4 stated, \textit{``Because most of it, when it's technical, it's, like, over my head ... not always extremely interesting in those meetings.}'' However, developers can be reluctant to take time to do this: \textit{``...they just don't really like the process''} (P1). This gap in technical expertise can lead to suboptimal evaluations, placing an additional burden on developers. P2 described their work as ``\textit{converting [domain knowledge] to technical expertise'' - since ``business people always are proposing different solutions as well, but technically, those weren't feasible or effective.}''

\subsubsection{KI2: Criteria inclusion depends on a hierarchical decision making process.} \label{KI2}
Across industry and academia, teams had well-defined hierarchies that dictated who was responsible for final decisions. While no single stakeholder was consistently dominant, a dominant stakeholder typically held the final say over the implementation of the criteria. For example, P1 described a scenario in which given requirements were not implemented, saying \textit{``they just picked some that they think the LLM could correctly do, and then they used it, and then definitely that's not capturing all the experience.''} This structure reflects a division of roles in which domain experts contribute requirements, but developers or project leads ultimately determine which criteria are implemented. While such hierarchies can streamline decision-making, they may also limit the integration of diverse perspectives, leading to criteria that do not fully capture stakeholders' expertise or account for uncertainties in AI outputs \cite{subramonyam2022solving}. To produce more robust and representative criteria, domain experts need to be meaningfully integrated into the decision-making process rather than consulted only at specific stages. In this sense, an explicitly defined decision-making structure can still support co-creation while allowing teams to converge on final criteria.

\subsubsection{KI3: Current workflows insufficiently support meaningful collaboration among domain experts} \label{KI3}
In our interviews, domain experts expressed a desire to consult with others to discuss the best way to handle specific topics or niche situations, something that was not always supported in existing workflows. For example, P10 stated \textit{``I think that I would want to do it (evaluation) with a group''}, while P6 emphasized the importance of their ability to say, \textit{``Let me go ask my colleagues.''} 

Though our participants emphasized the collaborative nature of knowledge work, they also identified some challenges in existing workflows. For example, though P10 wanted to work in a group, they also mentioned wanting \textit{``a little bit more time for like self-reflection first''}, similar to P9's request for \textit{``more time to process''} before talking to the group. Due to potential unfamiliarity with LLM-as-a-judge systems or criteria creation, domain experts may not be immediately ready to jump into a collaborative discussion, and conversations that occur too early in the familiarization process may lead to bias formation.
 
Therefore, criteria-creation workflows should account for collaboration among domain experts. This can ensure that criteria are more rigorous and reduce the burden of criteria drift. In addition, encouraging collaboration can also mitigate the effect of personal differences in criteria, as personal background and experience influence how criteria are expressed \cite{gebreegziabher2025metricmate}. Ensuring that human oversight comes from diverse sources is crucial to creating an equitable LLM judge.

\subsubsection{KI4: Human-AI alignment is constrained by development experts} \label{KI4}
While recent studies in LLM-as-a-judge assume a single user interacting with an AI system, our findings show that criteria creation in practice involves interdependent roles across multiple stakeholders \cite{Terry_Kulkarni_Wattenberg_Dixon_Morris_2024, Gu_Jiang_Shi_Tan_Zhai_Xu_Li_Shen_Ma_Liu_et}. This process splits goal articulation and output assessment across different roles. On one hand, domain experts evaluate responses from an LLM judge but may lack the technical expertise to directly steer or manipulate the system's LLM. On the other hand, developers can interact with the LLM but require requirements derived from domain experts. 

As a result, alignment is not solely a human-AI problem, but a \textit{human-human-AI} alignment problem, in which stakeholders must first align with each other before effectively aligning the system. Misalignment between stakeholders can lead to incomplete or distorted criteria, even when each individual perspective is valid.

One team we interviewed used the ``show-and-tell'' method, in which domain experts demonstrated the appropriate behavior, often paired with an incorrect judgment or response in the evaluation data. For example, in evaluating a chatbot, a domain expert may review an incorrect response and correct it, producing a correct response for the same input data. This information is then given to developers as a pair of positive and negative examples, which the LLM-judge should be trained to distinguish between. However, this process places the burden on developers to interpret implicit requirements and translate them into executable criteria, often without direct access to the underlying rationale.

\subsection{Design Goals}
Based on the insights from the formative study, we derive the following design goals:

\begin{enumerate}
    \item \textbf{DG1}: Support multiple stakeholders' alignment of criteria to more holistically facilitate the domain human-human-AI alignment process (Sections \ref{KI1}, \ref{KI4})
    \item \textbf{DG2}: Support an asynchronous co-creation workflow that enables broad participation in criteria development, while still accommodating hierarchical decision making for final decisions (Section \ref{KI2})
    \item \textbf{DG3}: Highlight and help users understand nuanced criteria and confidence levels, to avoid oversimplification of criteria requirements (Section \ref{KI3})
\end{enumerate}
\section{The \toolname{} System}

We implemented \toolname{}, a multi-user web-based system that enables different stakeholders to define, iterate on, and build collective LLM evaluation criteria for the LLM-as-a-judge framework~\cite{Gu_Jiang_Shi_Tan_Zhai_Xu_Li_Shen_Ma_Liu_et}. We draw upon Briggs’ Consensus Building Theory (CBT)~\cite{briggs2005toward}, which frames consensus building as a cyclical process: articulating a proposal, evaluating commitment, discovering and diagnosing conflicts, and resolving disagreements through discussion, potentially leading to revised proposals. CBT allows us to directly address \textbf{DG1} in supporting the consensus creation process, and provides the theoretical framework around which we build our system in order to support \textbf{DG2} and \textbf{DG3}. In our context, proposals correspond to criteria or criterion revisions, and disagreements may arise during their evaluation. CBT identifies several types of disagreement, which we adapt to the criteria-iteration process: 
\begin{itemize}
    \item \textit{Differences of Meaning}: ambiguity in the wording of criteria,
    \item \textit{Differences of Mental Model}: differing expectations of how criteria operate,
    \item \textit{Differences of Information}: an imbalance of information or knowledge, such as data and expertise, between evaluators, 
    \item \textit{Differences of Goals}: disagreement over desired outcomes, and 
    \item \textit{Differences of Taste}: differing value priorities.
\end{itemize}

  \begin{figure*}[htbp]
  \centering
  \includegraphics[width=\textwidth]{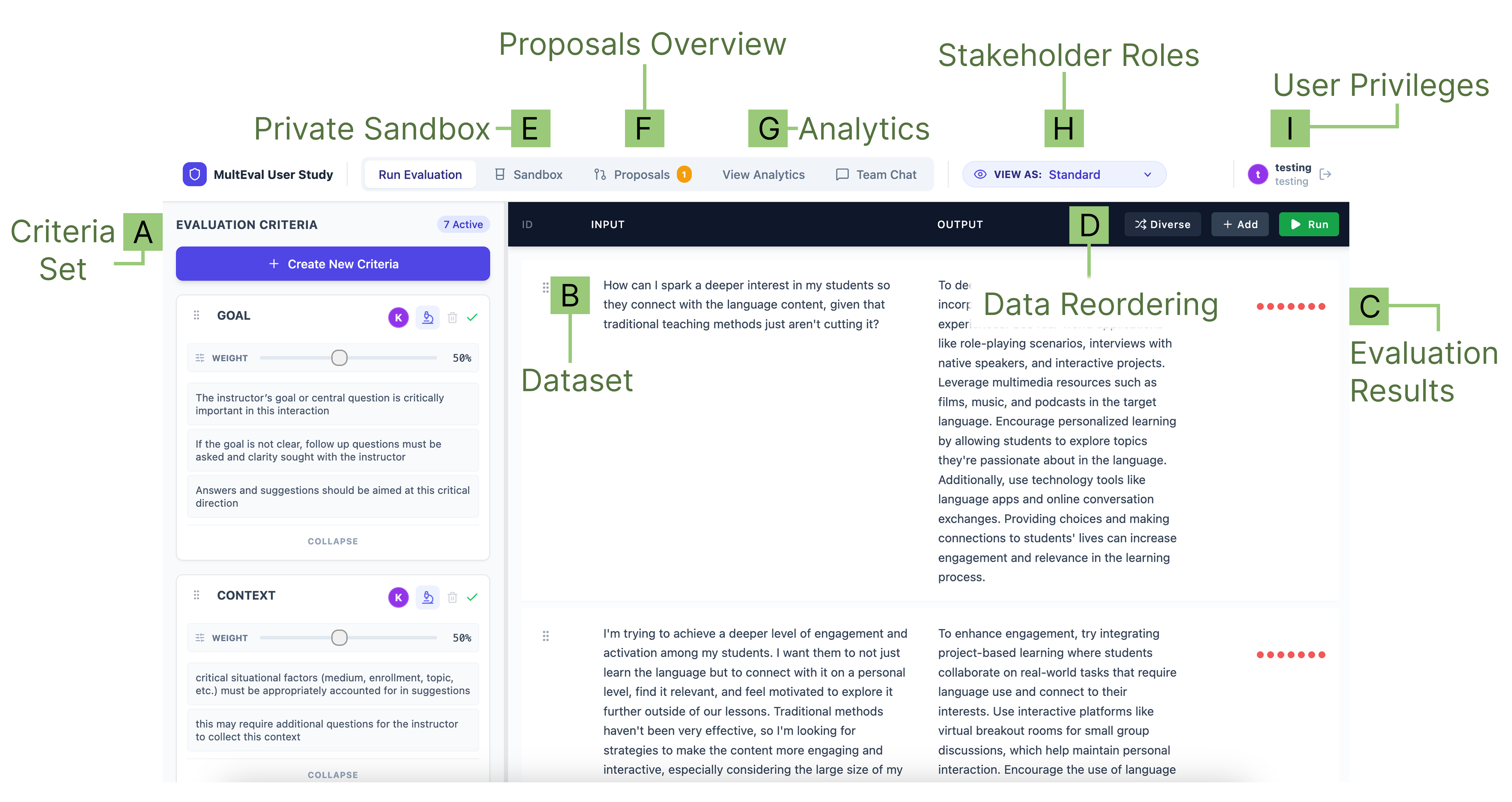}
  \caption{Overview of the \toolname{} System: A) The global criteria set, showing assertions, authorship, and criteria weight. B) The project's dataset, with input and output pairs. C) The results of the evaluation - red circles mean a failed criterion, while green means a pass. D) The option to reorder the dataset to show a diverse subset of data first. This feature helps surface disagreements faster. E) The private sandbox, where evaluators can test and author new criteria or propose changes to criteria. F) The proposal tab, showing active proposals as well as support from other evaluators and context to inform the changes. G) The analytics tab, where evaluators can see how criteria have changed throughout the process, as well as an overview of the contribution. H) A lightweight stakeholder-role swap, allowing users to quickly sample evaluation through a different lens. I) The user within the project has specific roles according to their background and expertise contribution - this gives additional context to authorship. Privileged `admin' users have final say on criteria inclusion.}
  \Description{Overview of the \toolname{} System}
    \label{fig:evaluation_screen}
\end{figure*}

A key step in consensus building is proposal evaluation. Here, \toolname{} applies criteria to data, producing outputs that serve as a shared evaluation artifact. When criteria are easily operationalized, disagreements about how they should be applied are resolved through execution; remaining conflicts instead reflect differing expectations for how the LLM-judge should interpret the criteria.

\subsection{Example Scenario}
Alice, Bob, and Charlie are the developer, domain expert, and project manager, respectively, of a new LLM-as-a-judge system. They decide to use \toolname{} to help them create the system's criteria. Charlie creates a project, naming it and uploading their dataset. Charlie adds Alice and Bob as evaluators, keeping admin privilege for themselves. In their initial meeting, the team has developed a basic set of criteria, which Charlie inputs into \toolname{}. Now, on their own time, Alice and Bob can explore the criteria set.

\begin{figure*}[!htbp]
  \centering
  \includegraphics[width=0.8\textwidth]{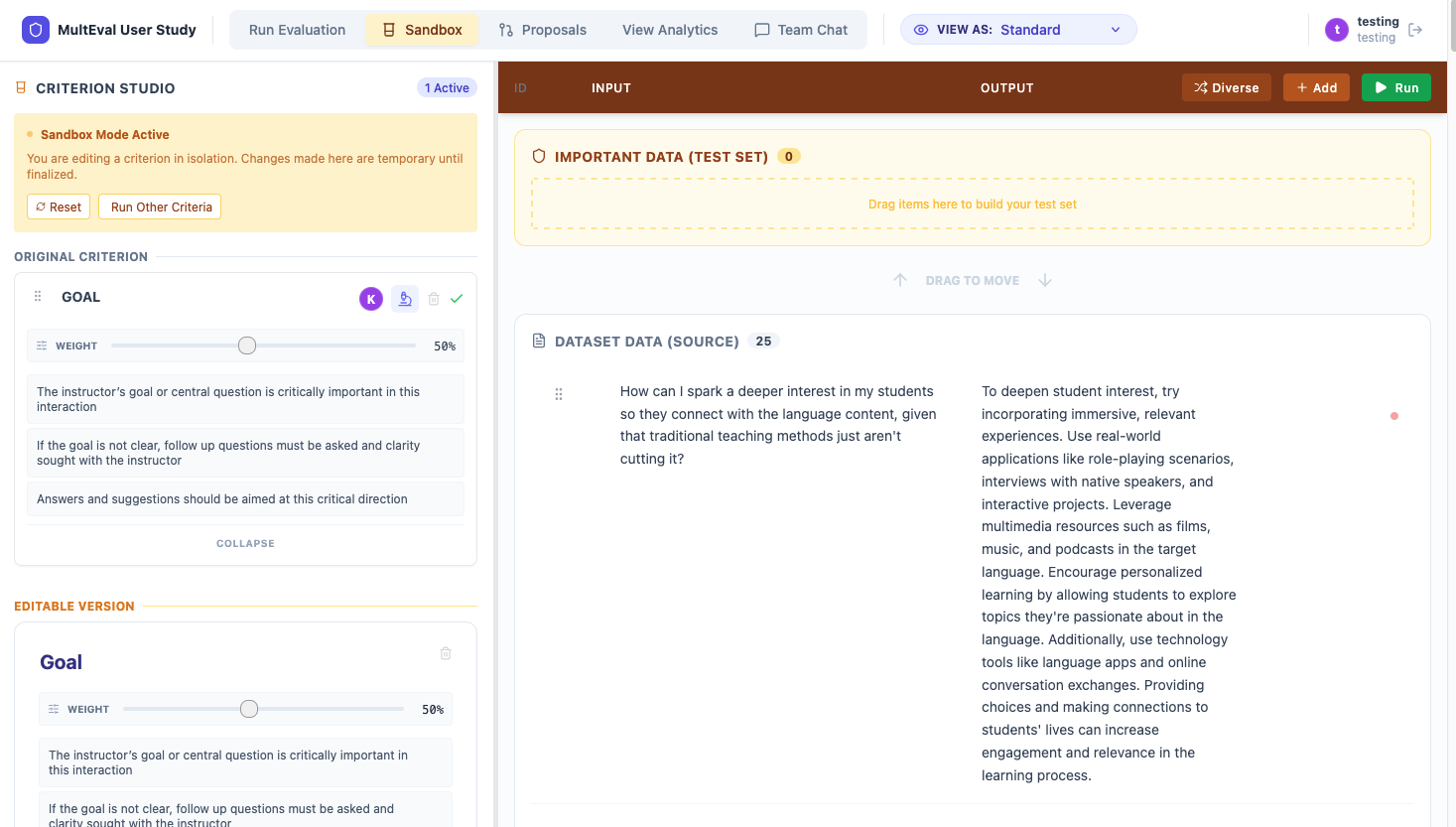}
  \caption{
    Private Sandbox. Users see the original criteria alongside an editable version on the left side panel. Users can additionally add data to a 'test set' or attach it to the criteria as context.
  }
  \label{fig:sandbox}
\end{figure*}

Alice sees criteria alongside the dataset and evaluation results (Figs. \ref{fig:evaluation_screen}, \ref{fig:tracing}). Scrolling through, Alice sees a criterion that has interesting wording. She selects it, taking her to her private testing sandbox (Fig. \ref{fig:sandbox}). Here, Alice further tests the criterion, noticing some data points that exhibit unintended behavior. She changes some words in the criteria and retests. Satisfied, Alice attaches the data point to the proposed change and submits the proposal for review.

Bob sees Alice's proposal and takes it to the sandbox dashboard for testing. He is not entirely confident about the specific wording that Alice uses, so Bob highlights a section of the proposed criteria. \toolname{} suggests a couple of different variations. Bob selects one of the variations and, testing the changes, he creates a new data point to highlight the difference. After some tweaking, Bob is satisfied with the changes and submits the proposal. Continuing to review other proposals, he shows support for some proposals by adding his vote. Later, Charlie reviews the proposals, starting with the ones that were voted on, and accepts or rejects them as he sees fit (Fig. \ref{fig:proposal-admin}). Over time, the criteria evolve as the team creates, reviews, and accepts proposals until they eventually converge. 

\subsection{Key Features}


\subsubsection{Criteria Proposal Loop}
\begin{figure}[!htbp]
  \centering
  \includegraphics[width=0.8\columnwidth]{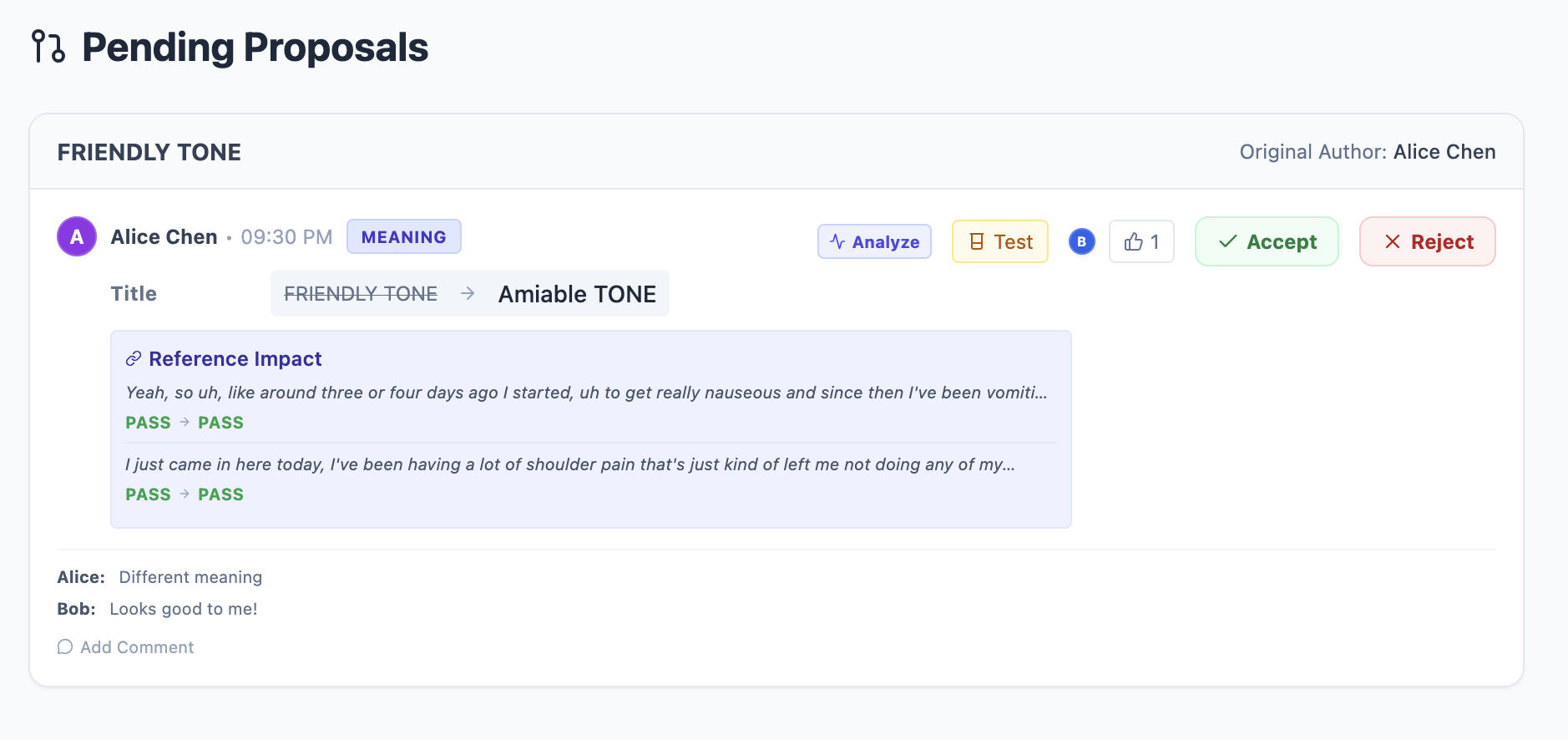}
  \caption{
    A proposal as seen by a user with administrative privileges. Admin users can accept or reject proposals. Non-admin users can only vote and comment on proposals.
  }
  \label{fig:proposal-admin}
\end{figure}

The application of CBT requires that \toolname{} makes disagreements explicit and actionable within the criteria development process. In \toolname{}, this involves surfacing differences not only between evaluators and the LLM judge, but also among evaluators themselves, while providing a private space for exploring and refining criteria. We implemented this as a private sandbox for individual evaluators to iterate on criteria, as seen in Fig. \ref{fig:sandbox}. Especially for evaluators without a technical background, creating a private space for exploration may help them become more familiar with the evaluation process. The sandbox also features a toggle that enables the remaining criteria, allowing users to check for any potential conflicts. Similarly, users must be able to propose modifications to criteria, along with some context for the change, as in \textit{Git} or other version control systems (Fig. \ref{fig:proposal-admin}). However, this context goes beyond a simple description of the change and its rationale. Since criteria are meant to dictate the LLM-judge's behavior, the context should include some measure of how the LLM judge's behavior has changed. For example, some data on which the change to the criteria has altered the result of the judgment. This feature supports \textbf{DG2} and \textbf{DG3}, allowing evaluators to understand the nuances within the evaluation of criteria at their own pace.

\subsubsection{Supporting Disagreement Diagnosis}
\begin{figure}[!htbp]
  \centering
  \includegraphics[width=0.7\columnwidth]{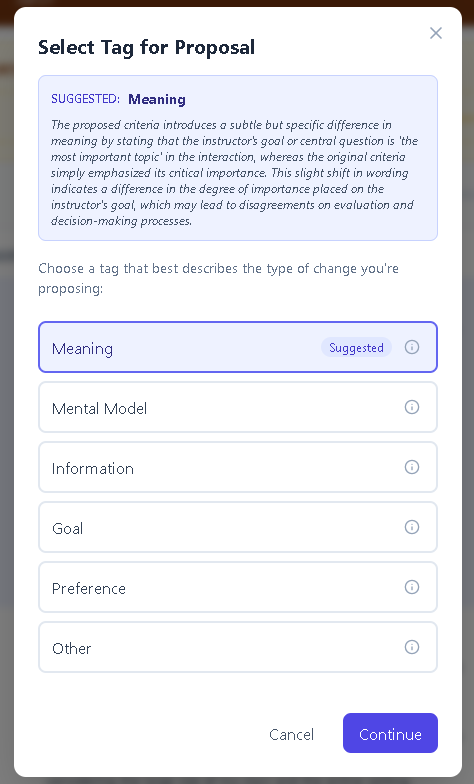}
  \caption{
    The system diagnoses the type of disagreement between the initial version of the criteria and the proposal. If the user disagrees with the rationale, they can select a different tag. Each tag shows a definition of its type of disagreement when the information icon is hovered over.
  }
  \label{fig:proposal-tag}
\end{figure}

When it comes to disagreements about criteria, surfacing which type of disagreement is the first step towards resolving the conflict. \toolname{} employs strategies to help users identify and classify disagreements. By determining which type of feature the user explored in their criteria iteration, \toolname{} can implicitly infer what the type of disagreement is. One such feature is the support for exploring different versions of highlighted sections of criteria. This allows users to iterate more quickly and suggests that the disagreement with the original criteria was related to meaning or goal. Similarly, by allowing users to attach data and author data, the system infers that the user may disagree with the information or have a mental model, respectively (Fig. \ref{fig:proposal-tag}). When users view proposals, they can `like' or `dislike' them to show their support.

\subsubsection{Surfacing Disagreement Through Data Curation}
Criteria iteration is a time-consuming process, partly because evaluators must review a large amount of data to ensure the quality of their judgments. Many data points in a dataset may be similar, forcing evaluators to filter through large quantities of data to find interesting results. To address this, \toolname{} implements an algorithm that identifies the most critical data that may result in disagreements. We first compute a fixed-length embedding using OpenAI's \texttt{text-embedding-3-small} model, then sorting using a greedy farthest-neighbors algorithm with cosine similarity. This ensures evaluators see structurally or semantically different examples rather than near-duplicate data, increasing the likelihood that differing interpretations of data appear earlier. In addition, once interesting data have been identified, \toolname{} allows users to select relevant data points and designate them as representative data. These data points are grouped and always tested, allowing users to see how their changes to the criteria affect specific, important data.

\subsubsection{Role-Cognizant Analysis}
To support transparency and reflection in the development of multi-stakeholder criteria, we implemented a version-control feature for evaluation criteria that explicitly surfaces contributors' roles and expertise over time (Fig. \ref{fig:version}). The system visualizes each criterion as a chronological timeline, beginning from its initial creation and progressing through successive versions to its current or finalized form. Each revision is annotated with metadata indicating the author's role, enabling users to contextualize changes through a lens of expertise and responsibility. By visualizing role-aware iterations, the timeline allows teams to identify where and how evaluators with different roles diverge in their judgments, priorities, or interpretations of quality. In addition, to encourage collaboration in criteria evaluation, \toolname{} offers lightweight stakeholder personas, based on the evaluator roles defined in the project. These personas take on different evaluator roles and background information, prompting the LLM-judge to adopt a similar persona in their evaluation. This feature aims to enable stakeholders to incorporate other perspectives at a low cost and with minimal time commitment. This feature was designed not as a replacement for real evaluators, but as a reminder for users to consult real evaluators.

\subsection{Implementation Details}
\toolname{} was created using a Node.js\footnote{\url{https://nodejs.org/en}} backend and React\footnote{\url{http://react.dev/}} frontend. For the case study, the system was deployed online, using Firebase\footnote{\url{https://firebase.google.com/}}. Specifically, the database and hosting were done using Firebase Firestore and Firebase Cloud Functions. For the LLM functions, we used OpenAI's GPT-5\footnote{\url{https://openai.com/}}. Prompts can be found in the Appendix \ref{appendix:system_prompt}.

\section{Case Study}
To evaluate our system, we recruited a real-world team of four pedagogy experts who had prior experience collaboratively evaluating a teaching assistant chatbot, and observed their interactions with \toolname{}. The study lasted 1.5 hours and was conducted online. During the study, the team of experts was introduced to the \toolname{} system and asked to complete a criteria creation task in the pedagogy domain. The study concluded with a semi-structured focus group-style interview.

\subsection{Participants}
We recruited a team of four pedagogical experts affiliated with the University of Notre Dame, a U.S. higher education institution. This team was identified due to their previous work, and invited via targeted email outreach based on its members' expertise in pedagogy, instructional design, and educational evaluation. All team members provided informed consent prior to participation and were compensated \$80. The study was approved by the University of Notre Dame's IRB.

The team had prior experience evaluating pedagogical chatbots in research settings. In that context, its team members collaboratively developed evaluation criteria and applied them to assess model outputs, gaining hands-on experience articulating quality dimensions for educational dialogue systems. We intentionally recruited this team to ensure they were comfortable both setting evaluation criteria and critiquing LLM-generated responses. All the team members held doctoral degrees and reported substantial professional experience (7--14+ years). The team had complementary perspectives spanning faculty development and instructional consulting, academic standards and assessment, writing pedagogy and the use of generative AI in instruction, alternative assessment methodologies, and equity-minded pedagogy. All reported prior use of LLMs in their professional workflows (e.g., teaching, research, instructional planning, and analytical reasoning).

\subsection{Procedure}
The team joined a Zoom meeting where they were introduced to the researchers and the study. All team members were asked to give brief introductions to one another. Participants were then introduced to the task and the dataset and given 10 minutes to individually create an initial set of criteria in a text editor. They were given a dataset of input/output pairs of instructor-related questions and responses to familiarize themselves with the data. Afterwards, the team was given a brief walkthrough of the \toolname{} system, and then had 30 minutes to use \toolname{} and collaboratively create criteria. During this time, team members were free to discuss during the Zoom call and had access to their initial criteria document. After the time elapsed, the researchers asked team members a series of questions in a semi-structured focus group interview format for 30 minutes.

\subsection{Dataset}
Because the studied team had pedagogical experts in academia, we asked them to review model outputs responding to instructor-authored pedagogical questions. The scenarios consisted of questions an instructor might pose to a pedagogy expert when reflecting on their teaching practice, such as how to approach instructional decisions, interpret student understanding, or respond to classroom challenges.

We used a set of previously developed pedagogical conversation questions and generated a new dataset of 26 single-turn model responses using OpenAI's GPT-4o model. Each response was generated using a teaching-assistant prompt that framed the model as a pedagogy-focused advisor, tasked with providing concise, expert-oriented guidance to instructors, as shown in Appendix \ref{appendix:synthetic_data}. These outputs were intended to approximate a baseline pedagogical consultation LLM chatbot.

\subsection{Qualitative Analysis}
We analyzed the focus group data using a collaborative thematic analysis approach \cite{braun2006using}. First, the focus group session was transcribed to capture both the discussion during use of the tool and the interview questions. Three of the authors then collectively reviewed the transcripts and imported relevant excerpts into a shared Miro \footnote{\url{https://miro.com/}} board. Using standard open coding procedures~\cite{williams2019art}, the authors read the excerpts and grouped segments that reflected similar ideas, experiences, or concerns. Through a series of iterative discussions, these groupings were refined, merged, or split as needed, resulting in higher-level themes. This process occurred over multiple rounds, with themes and labels evolving as the team revisited the data and negotiated shared interpretations. Any disagreements between coders were resolved through discussion until a unanimous agreement was reached. The final set of themes reflects consensus across the research team and captures recurring patterns observed across focus groups. We identified five themes, which we discuss in the next section.

\section{Evaluation of Automated Coding}
To assess whether \toolname{}'s LLM annotator pipeline could reliably apply the proposed taxonomy of expert disagreement, we conducted a validation study on a corpus of criterion-change proposals.

\paragraph{Reference Dataset.} We constructed a corpus of 40 criterion-change proposals drawn from a patient-intake chatbot domain. These proposals were generated using OpenAI's \textit{gpt-5} in a four-stage process. We began by defining the task as creating criteria for a baseline mental health consultation LLM chatbot. For each criterion, we generated a proposed alternative. Then, we randomly selected 10 criterion-change proposals and generated the reasoning behind the proposal. We repeated this step to generate the illustrative examples as well. In total, each proposal comprised an original criterion and a proposed alternative, with optional fields for the proposer's reasoning and an illustrative example. 

Two members of the research team independently coded each proposal across the five dimensions of the taxonomy. Initial inter-coder agreement, measured with Krippendorff's $\alpha$ at the nominal level, varied substantially across categories: Meaning ($\alpha = 0.48$), Goals ($\alpha = 0.41$), Mental Models ($\alpha = 0.07$), Information ($\alpha = 0.05$), and Taste ($\alpha = -0.09$). The coders then convened to adjudicate disagreements, discussing each divergent case until reaching consensus. This process yielded the consensus labels used as ground truth and informed iterative refinements to the codebook, including clarifications to category definitions and the addition of worked examples illustrating boundary cases.

\paragraph{Automated Annotation.} Automated annotation was performed with OpenAI's \textit{gpt-5.4-mini}. To isolate the contribution of different prompt components, we conducted an ablation study with three conditions. The \textit{baseline} condition presented the LLM with only the category names and the observation. The \textit{definitions-only} condition added the full codebook definitions for each category. Lastly, the \textit{full-prompt} condition further added worked examples illustrating positive, boundary, and cross-category cases. In all conditions, the LLM produced binary classifications and short justifications for each of the five categories.

\paragraph{Reliability Analysis.} Agreement between the automated annotations and the consensus ground truth was assessed using Krippendorff's $\alpha$ at the nominal level, computed independently for each category and each ablation condition. The full-prompt condition yielded the highest agreement on four of the five categories, indicating that worked examples contributed meaningfully to the LLM's calibration beyond what definitions alone provided, even though the \textit{Differences of Meaning} category decreased. Agreement varied substantially across categories, ranging from $\alpha = 0.605$ for \textit{Differences of Information} to $\alpha = 0.225$ for \textit{Differences of Meaning}. More details are reported in Appendix \ref{appendix:technical_evaluation}.

We interpret these results as a positive validation of \toolname{}'s annotation pipeline. The improvement from baseline to the definitions-only condition, and the further gains from adding worked examples, indicate that the codebook's structure and worked examples function as intended in guiding the model toward category-aligned classifications. The pattern across categories was consistent between the LLM annotator and human coders. The categories requiring more interpretive judgment showed lower agreement across both rater types. This consistency suggests that the pipeline tracks the analytical structure of the taxonomy rather than introducing model-specific artifacts, providing users with a concrete starting point for deliberation.
\section{Results}
\subsection{Starting with Common Ground and Organization}
Before using \toolname{}, the team established shared ground by pooling initial criteria and aligning on vocabulary. This early alignment appeared to reduce later friction during criteria authoring. This behavior aligns with the notion of establishing common ground within the team~\cite{clark1991grounding}. After participants spent a few minutes creating their own initial set of criteria, they were tasked with consolidating them into a unified set. When the group-criteria session began, participants immediately suggested consolidating their criteria sets, expecting to find overlap. P3 stated, \textit{``I do wonder if a starting place...to actually go back to each of our individual criteria, just sort of read through as a group what each of us put down, to get a sense of how we're all thinking about these criteria...?''}. This interaction was enabled by the session's synchronous nature, which allowed participants to talk to each other and coordinate. For instance, P3 mentioned that it was \textit{``hugely important just for grounding and understanding and thinking about next steps.''}. 

Although the team had access to \toolname{} from the beginning, they decided to first discuss their initial criteria for 10 minutes. During this time, participants also agreed on establishing a shared vocabulary to guide their discussion: \textit{``I just think it would be really helpful for us to have a common vocabulary as we discuss.''}~(P3). Despite these conversations outside of the system, this initial organization enabled the team to use \toolname{} more efficiently. P4 reported, \textit{``...because of the way that we initially set up our work ... we had a plan, we were organized... I think that organization really, really made our work more efficient in the tool''}~(P4). This also saved the administrator (P4) time on reviewing initial criteria since duplicates were removed: \textit{``...once somebody submitted a proposal, I just accepted it. There was no review that was necessary at that point''}~(P4).

\subsection{Struggle with Specificity}
Participants struggled to determine the appropriate level of detail and specificity when formulating criteria, often treating specificity as a primary lever for improving the judge's behavior. Beyond establishing a shared vocabulary, word choice remained a recurring challenge, as participants attempted to translate a high-level intent---what they wanted the judge to reward or penalize---into text they expected \toolname{} to apply consistently. For example, P3 noted uncertainty about phrasing: \textit{``I don't know if that is the word that is capturing what I'm trying to suggest there''}. Similarly, P2 emphasized a desire for criteria that provided stronger guidance by spelling out expectations more concretely, including through examples: \textit{``Yeah, I think we all liked the idea of specific or concrete examples... but I wanted more... I wanted to kind of spell it out for me even more.''}

This struggle reflects a broader sense-making problem in criteria authoring, wherein teams must determine whether evaluation failures are attributable to ambiguous language, missing constraints, or genuine differences in interpretation. In practice, this often meant negotiating what level of detail would be ``specific enough'' without becoming overly narrow. While participants did not escalate this into overt disagreement during the session, their comments suggest a latent trade-off between flexible writing criteria that span contexts and precise criteria that reduce divergent interpretations.

Participants also reflected on how the formatting of the criteria shaped their interpretation of the overall criteria set. For instance, P3 asked, \textit{``Do I need to be phrasing this in a certain way?''} This question highlights that, beyond the content of the criteria, participants were uncertain about the conventions for encoding them for an LLM judge. For example, they questioned how strongly to phrase requirements, whether to include examples, and how to structure the text. Because the study was time-limited, the team did not systematically test how such formatting choices affected \toolname{}'s evaluations. Nevertheless, some team members expressed skepticism that minor wording changes would meaningfully affect the tool's behavior. As P4 declared: \textit{``I suspect that it doesn't even make a difference for the actual application of the tool.''}, pointing to a need for tool support that makes the relationship between the criteria formulation and evaluation outcomes more transparent.

\subsection{Multiple Perspectives and Group Work}
Participants emphasized that criteria creation benefits from incorporating multiple perspectives, particularly different forms of expertise. For example, P1 argued for grounding criteria in shared evidence rather than individual judgment: \textit{``I'd appeal to the literature and the evidence, and so it's not just grounded in, you know, someone's personal experience''} ~(P1). 

Although all participants were experts in the task domain, they still recognized limitations in relying on a single perspective, suggesting that additional viewpoints could help capture nuances even within the same domain. P3 suggested support for \textit{``different personas... [to] capture maybe some of those discipline-specific contexts... and filter the results.''}. This highlights a perceived need to extend beyond a single expert perspective, even within a domain-aligned group.  

Participants also described collaboration as beneficial for both efficiency and ideation. They noted that working together was substantially faster than if they had been working individually (\textit{``...it just would have taken much, much longer [individually]}''~(P3)) and enabled a broader set of criteria (\textit{``...instead of 7 criteria that we all came up with together, you'd have probably had two or three...''}~(P1)). An additional benefit was the ability to specialize. As P1 noted, \textit{``in a group ... we're all contributing from our different areas of expertise and playing to our different strengths and capacities''}, which supported both faster progress and more in-depth analysis.

There were a few disagreements in the group. Some participants raised minor concerns about the formatting of particular criteria. However, participants generally reached agreement quickly during the session. For example, one small disagreement centered on naming conventions: \textit{``What is the goal of what the naming of the criteria should do? Is it supposed to be descriptive and be able to communicate to the system, or is it necessarily about uniformity, and how do we balance those?''}~(P1). This was largely resolved through discussion, and the group ultimately decided not to pursue it further because they believed it would have little impact on the final output of the criteria set.

Participants noted that they would have been more likely to disagree if the criteria produced unintended behavior or conflicted with their personal experience. For example, P2 remarked that they would have argued more strongly for a criterion tied to their own experience (\textit{``the suggestion I made about making sure it's answering the right question ... I feel like I would have been more confident in arguing for it... because it was something that I had personal experience with''}~(P2)). Similarly, P3 suggested they would object if a proposed criterion was \textit{``deeply against my own pedagogical philosophy''}~(P3). In these cases, participants emphasized that resolving disagreements may require surfacing the rationale behind strongly held beliefs rather than appealing to authority; as P4 reflected, \textit{``...it may just come down to... damn it, just trust me on this, which is not a good argument... kind of the why am I believing this and believing it so strongly?''}

\subsection{Negotiating Roles and Hierarchy} 
The team members initially resisted hierarchy. At the beginning of the study, participants were asked, as a group, whether they would designate a single teammate as an administrator. As coordination costs became more salient, they converged on an admin role. However, the role's authority remained contested.  The admin had the privilege to accept and reject proposals and delete criteria. The group was initially hesitant to designate an administrator: \textit{``we're all equal''}~(P4). However, some participants wished to designate P4 \textit{``I was going to nominate you, [P4].''}~(P3).

Later, shortly after the criteria creation session began, a few participants nominated P4 again: \textit{``I feel like I like the administrator idea more than I thought I did 10 minutes ago''}~(P2). The participants eventually agreed that P4 would take on the administrator role. The reason P4 was chosen was partially due to some participants having worked with them prior (``\textit{I've worked with [P4] a lot, so I know his strengths in note-taking and organization}''~(P1)), but also for P4's apparent familiarity with the task domain: \textit{``it seemed like P4 had a clear direction.''}~(P2). 

After the session, participants reflected further on why they did not want to take the admin role, attributing it to uncertainty about the task and the pressure of the role. For example, P2 said \textit{``I think mine stemmed from uncertainty...I did not want to be the administrator because I was nervous about not fully grasping what we were doing, and so I was like, let's just all throw our hats in the ring and see what emerges.''} Overall, participants' reluctance to take on the role suggests that administrative authority can feel high-stakes when the task and decision boundaries are not yet well defined.

In this case study, assigning an administrator did not fully resolve questions of authority. Instead, it revealed a mismatch in expectations about the role. Some participants framed the administrator as a decision-maker who would determine which criteria were included in the final set, which aligns with \toolname's permission model. In contrast, P4 described the role as primarily (\textit{``just a logistical administrative role, not ... the final arbiter of decisions.''}) This suggests that adopting lightweight hierarchies may help reduce coordination overhead, but teams may still benefit from mechanisms that make accountability explicit (e.g., who can accept changes, who can veto, and how disagreements are resolved).

\subsection{Making Sense of the Interface}
Overall, participants described \toolname{} as easy to learn and navigate, noting that the workflow felt intuitive once they understood the basic proposal loop (e.g., testing in the sandbox, submitting proposals, and reviewing changes). For instance, P3 declared that \textit{``I think the interface was pretty clear... I like the way it sort of helped you work through the workflow.''}. While we did not administer a standardized usability instrument, participants raised minor usability concerns in the post-task discussion, such as inconsistent behavior of text input fields (e.g., when edits were saved or reflected across views). As P4 noted, \textit{``Once I figured it out, it's not a big deal.''}

Additionally, P1 reported some difficulty understanding how certain features worked, particularly leaving comments on proposals: \textit{``I just wasn't sure if those comments would be... how would those comments work? Are they connected to the system and the information that we're giving it, or is it just for, like, people on the team to see, oh here's how [P1] was thinking''}~(P1). Beyond confusion about how features interfaced with the system, the feature set and terminology could be cognitively demanding: \textit{``that was a little bit confusing, and just the language, from new requirements to assertion to comments, how do those things interface''}~(P1). Overall, despite these issues, participants still considered the system generally usable with a clear workflow and interface.

\section{Discussion}
Our findings show that authoring criteria collaboratively is fundamentally a process of coordination and sense-making. In our case study, the team established common ground, negotiated shared vocabulary, and iterated on wording to manage ambiguity and specificity. We also found that \toolname{}'s collaboration features are most valuable when they preserve rationale and coordination context over time, enabling teams to revisit decisions and understand how the criteria evolve. In the following subsections, we discuss the key implications of these findings for the design of systems that support multi-stakeholder criteria creation.

\subsection{Supporting Continuity across Asynchronous and Synchronous Work}
Our findings suggest that the value of \toolname{} becomes most visible in settings where coordination cannot rely on real-time communication. In our study, participants used Zoom to externalize reasoning, negotiate wording, and resolve minor disagreements in real time. This synchronous channel reduced the need for in-system mechanisms to surface and negotiate conflicts. 

However, in many real-world settings, teams cannot rely on continuous synchronous interaction due to limited overlap in availability or the need for individual reflection. In such cases, the work of establishing shared vocabulary and negotiating criterion specificity---both central to our findings---must be carried through the system itself. Without support, these decisions are easily lost, fragmented, or repeatedly renegotiated.

Features such as comments on proposals, attached examples, and version history can support \textit{asynchronous grounding} by preserving both decisions and their justifications. These features allow team members who were not present at earlier stages to understand why the criteria evolved. Participants explicitly pointed to this potential, as P3 noted, \textit{``I didn't quite understand the comment section either, but once you explained it, I was like, oh, right, if we were doing this asynchronously, that would be a really useful feature there.''}

In practice, our findings suggest that collaborative criteria authoring often alternates between periods of real-time coordination and more distributed, time-shifted work. Rather than prescribing a specific workflow, this highlights the need for systems like \toolname{} to support continuity across these models. In particular, features such as proposals, comments, and revision history can help surface disagreements as they emerge and prevent them from being lost across fragmented communication. These mechanisms also enable contributors to quickly recover context after periods of absence, supporting smoother transitions between synchronous and asynchronous collaboration.

\subsection{Making Coordination Visible Through System Translucency}
Across the session, participants frequently relied on the external initial criteria document to coordinate work alongside \toolname{}, particularly to track task assignments (e.g., which criterion they were responsible for) and avoid redundant effort. This suggests that visibility into who is working on what---and how work is evolving---is a central need for discussing and negotiating criteria. 

We frame this need through the lens of social translucence, where systems support collaboration by making activity, intentions, and contributions visible to others~\cite{erickson2000social}. In this context, translucency enables coordination without requiring constant real-time communication. Features such as authorship, version history, proposals, and comments can serve as a coordination infrastructure, making ongoing work and decision-making processes legible over time.

Participants implicitly enacted this workflow when dividing labor and reviewing each other's contributions. As P3 described, \textit{``should we all assign ourselves one of these, begin trying to articulate and capture what [Team-Member] is, you know, helpfully putting down here, and then we can sign off on one another's, or make amendments as needed''}. However, the current version of \toolname{} provides limited support for making these collaboration signals explicit. For example, lightweight voting mechanisms (likes and dislikes) are underspecified: a ``like'' could indicate agreement or acknowledgment, while a ``dislike'' could indicate disagreement, uncertainty, or an alternative. 

These findings suggest a need to make social signals more interpretable and actionable. One direction is to clarify their semantics (e.g., separate buttons for \textit{reviewed}, \textit{support}, and \textit{block}) and pair them with a brief, structured rationale. More broadly, translucency could be strengthened with at-a-glance indicators that support both parallel and sequential workflows (e.g., per-criterion badges for current editor, last updated time, pending proposals, and unresolved comments). Such features would reduce coordination overhead by making responsibilities, progress, and outstanding decisions visible without requiring synchronous interaction \cite{erickson2000social}.

\subsection{Evolving Authority Human-Human-AI Alignment}
The tension surrounding the administrator role reflects a broader human–human–AI alignment challenge. As discussed in the formative study, stakeholders must first align with each other before they can effectively align the system. In our study, this dependency became visible through participants' hesitation to assign epistemic authority early in the process. While participants were willing to delegate procedural authority---such as managing proposals or merging changes---they resisted granting any individual the power to define criteria on behalf of the group while their shared understanding remained unstable. 


The hesitation stems from the dual role of criteria as both a coordinator artifact and the mechanism through which the AI is aligned. Assigning authority over criteria is therefore not just a matter of organizing teamwork, but of determining how the system will enact and evaluate outputs. As a result, disagreements were not only interpersonal, but also tied to uncertainty about downstream AI behavior. 

As participants converged on shared interpretations, they became more willing to consolidate authority in an administrator role. However, this authority emerged gradually and remained negotiated. This pattern suggests that systems supporting collaborative LLM evaluation should not require fixed roles upfront, but instead enable graduated authority that evolves alongside shared understanding and trust in the system. More broadly, the study highlights that human–human–AI alignment unfolds progressively, requiring systems to support not only shared understanding, but also the gradual consolidation of authority as teams become confident in how their criteria shape system behavior.


\subsection{Understanding How Criteria Shape LLM Behavior}
A recurring challenge in the case study was participants' limited ability to anticipate how changes to criteria wording would affect the LLM judge's behavior. Although \toolname{} made criteria revisions and their outcomes visible, participants often questioned whether specific edits---such as changes in phrasing, structure, or examples---had any meaningful impact. This uncertainty made it difficult for the team to assess progress or determine whether refinements were actually improving alignment. 

We interpret this as a lack of \textit{behavioral legibility}: the team did not have a clear mental model of how symbolic changes to criteria translate into the behavior of an opaque, probabilistic LLM evaluation process. As a result, even when participants identified issues in the criteria, they were hesitant to assert strong positions or exercise epistemic authority without confidence in downstream effects. In this way, limited behavioral legibility not only constrained human-AI alignment, but also slowed human-human coordination.

Our findings suggest that supporting collaborative criteria authoring requires more than tools for negotiation and versioning. Teams also need mechanisms to reason about the sensitivity of criteria---understanding which changes matter, under what conditions, and why---so they can more confidently align their judgments with LLM behavior.

\subsection{Feature Usage}
Not all features were used during the session, and this non-use was informative rather than incidental. In many cases, it reflected a mismatch between the system's capabilities and participants' workflow, attention, and mental models.

First, several features were not easily discoverable within the flow of criteria authoring. When participants were focused on drafting and revising wording, they rarely explored other actions, suggesting that feature visibility alone is insufficient when attention is tightly coupled to the task at hand.

Second, participants' uncertainty about how \toolname{} interprets criteria shaped how they engaged with the system. In particular, doubts whether formatting and minor wording changes would meaningfully affect evaluation outcomes made it difficult to judge when to use features designed to support iteration and diagnosis. As a result, participants were less likely to engage in systematic exploration of the tool's capabilities.

Third, some features were implicitly designed for use across time, rather than within a single, time-bounded synchronous session. Artifacts such as comments, attached examples, and proposal history are most valuable when teams need to preserve rationale across contributors or revisit decisions later, which limited their perceived relevance during synchronous collaboration.

Taken together, these findings suggest that supporting collaborative criteria development is not only about providing functionality, but also about aligning features with moments of need. One design implication is to more tightly couple advanced features to breakdowns and decision points---for example, when a proposal is contested, when an evaluation appears surprising, or when a criterion is repeatedly revised. More broadly, improving the legibility of how criteria shape system behavior may help participants better anticipate when and how to engage with these features.

\subsection{Limitations and Future Work}
We acknowledge the main limitations of this study. First, our case study involved only one team of pedagogical experts with shared institutional and professional backgrounds, thereby limiting the generalizability of our findings. Rather than treating this as representative of all multi-stakeholder settings, we view this case as a \textit{best-case scenario} for collaborative criteria creation: participants shared domain expertise, professional norms, and a common evaluative ethos, reducing the likelihood of overt value conflict. Notably, even under these relatively aligned conditions, participants encountered substantial coordination and sense-making challenges, including negotiating specificity, interpreting system affordances, and clarifying decision authority. We therefore interpret these challenges as conservative indicators of the work required to collaboratively author evaluation criteria.

Second, the team composition is also not necessarily reflective of the diversity seen in real-world teams. While workplace teams may have more members and a wider range of roles, the team in our study was relatively small and uniform, with all four members sharing the same domain expertise and professional context. As P4 noted, \textit{``we've got a shared ethos... we belong to some of the same professional organizations... There's a lot of shared concerns, practices, there's a shared intellectual heritage that many of us have''}~(P4). More teams across different domains are needed to increase the study's sample size and to generalize its results and conclusions. In more heterogeneous teams—spanning domains, organizational roles, or value systems—we expect such tensions to be more pronounced, further motivating the need for tools that make disagreement, rationale, and coordination processes explicit. While \toolname{} was built to support diverse teams, the types of disagreements and the ability to resolve conflicts may vary across environments.
 
Third, the study's time limit was a drawback, as participants were unable to fully explore relevant system features in the allotted time. In the post-task interview, participants described challenges that may have been addressed or mitigated by features they did not use. Additionally, participants completed only a few criterion-iteration rounds due to limited time to evaluate the criteria they created. We anticipate that, with larger teams over a longer time period, \toolname{} may encounter scalability issues due to the number of proposals and criteria being created. Especially in high-conflict environments, additional work may be needed to reconcile individual differences.

Another limitation deals with the technical evaluation of our disagreement diagnosis feature. The automated annotator is not proposed as a substitute for human coding. Instead, it functions as an initial scaffold to support proposal authors and evaluators in initiating discussion and working toward consensus. Although the annotator may not produce perfectly accurate initial labels, it still surfaces potential disagreements that warrant further examination. Moreover, because the process of reaching consensus is often time-consuming and cognitively demanding, this feature is designed to reduce that burden by structuring and prompting more efficient deliberation. The role of human judgment remains a vital part of consensus behavior, and this feature exists to initiate and support those conversations.


Finally, the study setup could be enhanced to determine the specific benefits provided by \toolname{}. Rather than a case study where team members interact synchronously over Zoom, future work could use a deployment study in which teams collaborate both synchronously and asynchronously over a longer period.
\section{Conclusion}
We present \toolname{}, a multi-user, web-based system for collaboratively authoring and iteratively refining evaluation criteria for LLM-as-a-judge workflows. Grounded in consensus-building theory, \toolname{} supports teams in surfacing and diagnosing disagreement, preserving rationale through proposals and examples, and coordinating criteria evolution through role-aware history and review. Our case study suggests that collaborative criteria creation involves substantial coordination and sense-making work---from establishing shared vocabulary to negotiating decision rights---and that tools can reduce friction by making these processes visible and persistent, particularly in asynchronous settings.

More broadly, this work reframes evaluation in LLM-as-a-judge systems as a fundamentally social process. Rather than treating criteria as fixed technical specifications, we show that they are negotiated artifacts that emerge through interaction, uncertainty, and alignment among stakeholders. In this sense, evaluation is not solely a human–AI alignment problem, but a human–human–AI alignment challenge: teams must first converge on shared interpretations before those interpretations can be reliably encoded into system behavior. Designing for this reality requires moving beyond static evaluation pipelines toward systems that support evolving understanding, distributed expertise, and ongoing alignment across people and systems. We see this as a critical direction for building more transparent, accountable, and trustworthy AI evaluation practices.

\begin{acks}
This work was partially supported in part by the Notre Dame-IBM Technology Ethics Lab, an IBM Ph.D. Fellowship, Alfred P. Sloan Foundation G-2024-22427, the U.S. National Science Foundation under grant CNS-2426395, a Google Research Scholar Award, an NVIDIA Academic Hardware Grant, an Amazon Science Award, and a gift from Adobe Inc. 
\end{acks}

\section*{Generative AI Usage Disclosure}
The authors used ChatGPT and Gemini to assist in writing and proofreading the paper. The usage of generative AI in writing did not affect the intent or original meaning of the text. All words written were read and checked by a human author before submission. The authors also used Gemini and Cursor to assist in coding. The placement and inclusion of features were done solely by the authors. All decisions regarding code were made by the authors.

\bibliographystyle{ACM-Reference-Format}
\bibliography{bibliography}

\appendix
\section{System Prompts}
\label{appendix:system_prompt}

\subsection{LLM-Judge System}
\promptbox{\texttt{Judge Prompt: "You are a helpful assistant that can check the quality of an input-output pair based on a list of provided requirements. You can objectively evaluate the output on the given criteria based on its quality of responses by assessing how well the responses satisfy the given requirements in the criteria and assertions. The requirements will have examples that satisfy (pass) the requirements and ones that fail. You should provide comprehensive feedback on the responses according to each of these criteria and provide detailed justification for your feedback. If you refer to specific fragments of the responses in your feedback, you should also return these fragments as evidence. You should return your final answer as a valid JSON object of the following format. {"results": [{"assertion\_id": "<ID>","result": "<pass or fail>", "reason": "<comprehensive and detailed explanation of why the output does or does not satisfy the criterion>", "evidence": ["<maximum of 5 words or short phrases from the output that serve as evidence for your feedback>"]}]}"}}

\subsection{Rephrasing}
\promptbox{\texttt{Rephrasing prompt: "You are a helpful assistant that can paraphrase a short piece of text. Your response should be of similar length to the input text, within one or two words. You should not simply use synonyms, but rephrase the text in a way that keeps the same meaning. Your response should ONLY be the paraphrased text, nothing else."}}

\subsection{Synthetic Data Generation Prompts}
\label{appendix:synthetic_data}

\promptbox{\texttt{You are a pedagogy expert bot. You will introduce yourself as a teaching assistant bot whose role is to help the teacher with some questions about their teaching work. Your role is to provide specific advice and support to the teacher, one question at a time. Only ask follow up questions if your answer depends on necessary information that was not provided. Your response should be concise, within 100 words, addressing the specific question.}}

\subsection{Tag Suggestion}
\label{appendix:tag_suggestion}
\promptbox{\texttt{Tag suggestion prompt: You are an expert analyst of stakeholder disagreement and deliberation frameworks. Your task is to diagnose the *primary underlying source of disagreement* between two criteria statements. You must classify the disagreement using **exactly one** of the following tags, selected according to the *earliest applicable level* in the hierarchy below (i.e., start at 1 and only move to the next level if the previous one does not apply):1. **Difference of Meaning**   The same words, phrases, or symbols are interpreted as referring to different concepts or definitions.2. **Difference of Mental Model**   The meaning is shared, but there is a difference in beliefs about how the proposal functions, what mechanisms are involved, or what outcomes will result.3. **Difference of Information**   The disagreement arises from asymmetric, missing, or conflicting factual information, evidence, or assumptions about the world.4. **Difference of Goals**   The criteria reflect incompatible or competing objectives, even when meaning, mental model, and information are aligned. 5. **Difference of Taste**   The disagreement is rooted in normative preferences or value judgments about what *ought* to be prioritized, even when goals are shared. Instructions - Compare the **Original Criteria** and **Proposed Criteria**. - Consider the **Proposer's Comment** (if provided) to understand the proposer's intent and reasoning. - Consider the **Reference Examples** (if provided) to understand what specific data or scenarios motivated the change. - Determine the category that best explains the disagreement. - Do **not** select multiple categories. - Justify your decision by explicitly referencing how the two criteria differ, and incorporate insights from the comment and reference examples if they provide relevant context. - If multiple categories seem plausible, explain why the selected category is the most fundamental. Proposals with few word changes may be more likely to be differences of meaning, proposals with additional data attached may be more likely to be differences of mental model or information, and proposals with large changes may be differences of goals or values. \{Output Requirements\}}}

\subsection{Basic Tag Suggestion}
\label{appendix:basic_tag_suggestion}
\promptbox{\texttt{Classify this proposed change to the original criteria as either a difference of
Meaning, Mental Model, Information, Goals, or Taste. \{Output Requirements\}}}

\subsection{Output Requirements}

This prompt was appended to the tag suggestion prompts in order to generate the output in the correct format.

\promptbox{\texttt{Output Requirements - Respond **only** in valid JSON. - Follow this exact schema: {"tag": "<one of the five tag names>", "rationale": "<detailed explanation of why this tag was selected, referencing specific differences between the criteria>"} - Do not include additional commentary outside the JSON. Valid tag values are exactly: "Difference of Meaning", "Difference of Mental Model", "Difference of Information", "Difference of Goals", "Difference of Taste".}}

\section{Formative Study Participants}

\begin{table}[hbt]
\caption{Participant Attributes}
\label{tab:participants}
\small
\setlength{\tabcolsep}{5pt}
\renewcommand{\arraystretch}{1.12}
\begin{tabularx}{\textwidth}{@{} 
    >{\centering\arraybackslash}p{0.8cm}      
    >{\centering\arraybackslash}p{1.2cm}      
    p{1.8cm}                                  
    >{\raggedright\arraybackslash\hsize=0.8\hsize}X 
    >{\raggedright\arraybackslash}p{3.0cm}    
    >{\raggedright\arraybackslash\hsize=1.2\hsize}X 
@{}}
\toprule
\textbf{ID} & \textbf{Gender} & \textbf{Education} & \textbf{Primary Industry} & \textbf{Stakeholder Type} & \textbf{Evaluation Roles} \\
\midrule
P1  & F & Doctorate  & Education & Frontend Developer; Domain Expert & Modeling/Training; End-user Testing \\
P2  & F & Doctorate  & Software, Information Services, and Data Processing & Backend Developer; Domain Expert & Modeling/Training \\
P3  & M & Bachelors  & Telecommunications & Backend Developer & Modeling/Training; End-user Testing \\
P4  & M & Doctorate  & Education & Domain Expert & Domain Expert \\
P5  & M & Bachelors  & Software, Information Services, and Data Processing & Backend Developer & Testing/QA; Modeling/Training; End-user Testing \\
P6  & M & Doctorate  & Pedagogy Support & Domain Expert & Testing; Data Provision; Criteria Development \\
P7  & F & Doctorate  & Education Administration & Domain Expert & Design; Criteria Development \\
P8  & M & Doctorate  & Pedagogy Support & Domain Expert & Testing; Design; Data Provision; Criteria Development \\
P9  & F & Doctorate  & Pedagogy & Domain Expert & Testing; Criteria Development \\
P10 & M & Doctorate  & Education & Domain Expert & Testing; Criteria Development \\
\bottomrule
\end{tabularx}
\end{table}

\section{Technical Evaluation}
\label{appendix:technical_evaluation}

\begin{table}[hbt]
\caption{Krippendorff's Alpha scores. Greatest scores are bolded.}
\begin{tabular}{llllll}
Differences of... & Meaning & Mental Model & Information & Goal & Taste \\ \hline
Baseline                 & -0.580 & -0.049 & 0.428 & -0.159 & 0.134 \\
+Definitions             & \textbf{0.433}  & 0.393  & 0.604 & -0.053 & 0.172 \\
+Definitions \& Examples & 0.225  & \textbf{0.540}  & \textbf{0.605} & \textbf{0.263}  & \textbf{0.280}
\end{tabular}
\end{table}

\section{Additional System Images}

\begin{figure*}[!htbp]
  \centering
  \includegraphics[scale=0.4]{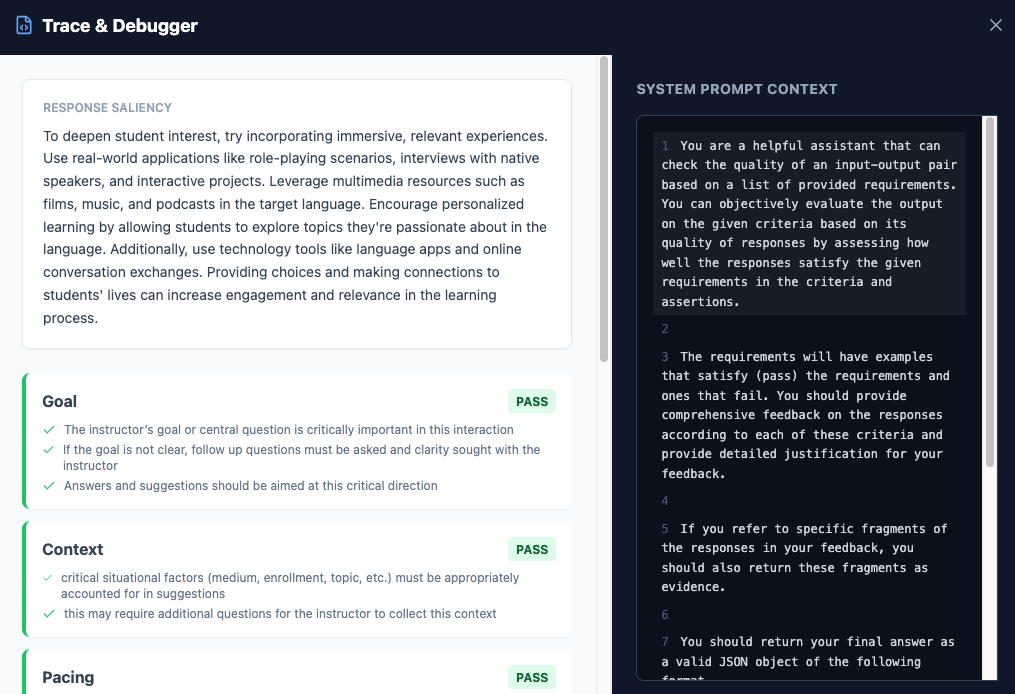}
  \caption{
  Example output of a data point trace and prompt.
  }
  \label{fig:tracing}
\end{figure*}

\begin{figure}[!htbp]
  \centering
  \includegraphics[width=\columnwidth]{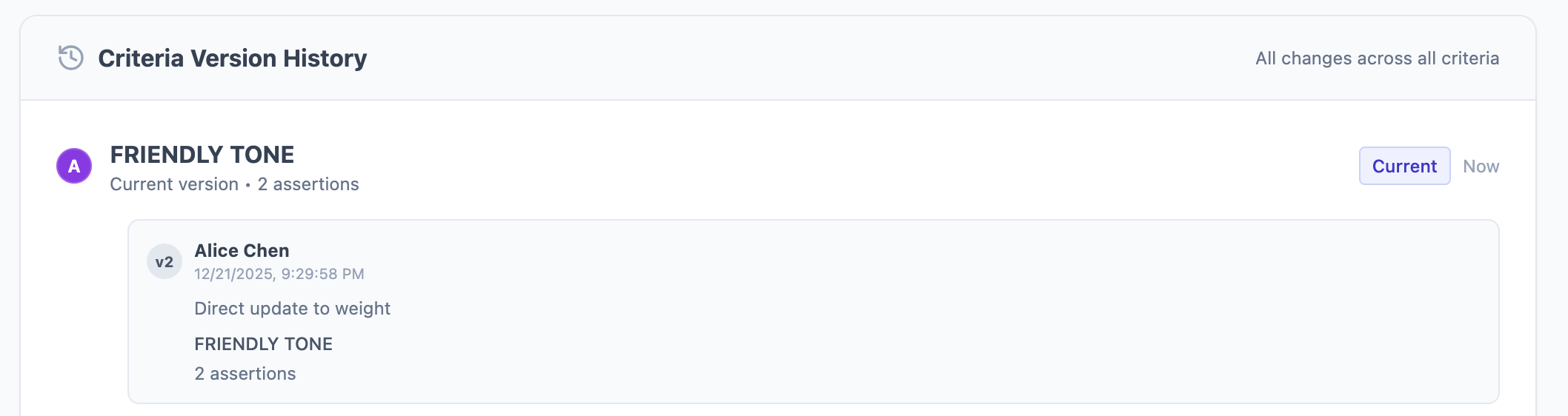}
  \caption{
    Timeline showing different versions of a criterion, including proposals.
  }
  \label{fig:version}
\end{figure}

\end{document}